\documentclass[twocolumn,showpacs,groupeaddress,preprintnumbers,amsmath,amssymb]{revtex4-1}
\usepackage{graphicx}
\usepackage{dcolumn}
\usepackage{bm}
\usepackage{hyperref}
\usepackage{epstopdf}
\usepackage{color}

\begin{document}
\title{Strange nonchaos in self-excited singing flames} 
\author{D. Premraj$^{1}$, Samadhan A. Pawar$^1$, Lipika Kabiraj$^2$, R. I. Sujith$^1$}
\address{$^1$Indian Institute of Technology Madras, Chennai 600 036, India\\
$^2$Indian Institute of Technology Ropar, Punjab 140 001, India}
\received{: to be included by reviewer}
\date{\today}
\begin{abstract}

\par We report the first experimental evidence of strange nonchaotic attractor (SNA) in the natural dynamics of a self-excited laboratory-scale system. In the previous experimental studies, the birth of SNA was observed in quasiperiodically forced systems; however, such an evidence of SNA in an autonomous laboratory system is yet to be reported. We discover the presence of SNA in between the attractors of quasiperiodicity and chaos through a fractalization route in a laboratory thermoacoustic system. The observed dynamical transitions from order to chaos via SNA is confirmed through various nonlinear characterization methods prescribed for the detection of SNA.
\end{abstract}

\maketitle 


\par Coupled nonlinear systems exhibit various kinds of dynamical behaviours including periodic, quasiperiodic, and chaotic oscillations \cite{ml, feud}.  Among these dynamics, one of the commonly observed state in quasiperiodically driven nonlinear systems is a strange nonchaos. Although strange nonchaotic attractors (SNAs) show similarity to chaotic attractors by having a fractal geometrical structure, SNAs are insensitive to initial conditions unlike the chaotic attractors \cite{Prasad1}. Grebogi {\em et al.} \cite{greb} was the first to report the possibility of SNAs in the system of quasiperiodically forced map. Afterwards, several numerical studies have demonstrated the existence of SNAs in systems such as pendulum \cite{pend1}, Duffing oscillator \cite{duff1}, logistic map \cite{log1}, Henon map \cite{henon}, and circular map \cite{circular}.

\par The experimental discovery of SNA was reported by Ditto {\em et al.} \cite{ditt} in a quasiperiodically forced system with a buckled magnetoelastic ribbon. In subsequent years, there have been several experimental observations of SNAs in practical systems \cite{Guan,kt,venk2,prem,paul}; however, all these studies presented the necessity of having quasiperiodic forcing to generate SNAs. Contrary to these studies, Negi {\em et al.} \cite{ram} showed theoretically that the need of quasiperiodic forcing is not mandatory for the creation of SNAs, and it could happen in naturally driven systems as well without the need of external forcing. Recently, Lindner {\em et al.} \cite{Lindner} showed the observation of SNAs in the natural system of a pulsating star KIC 5520878 network. However, to the best of our knowledge, there has not been a single experimental evidence of SNA reported in self-driven laboratory systems until now.

\par Most of the recent studies are focused on identifying the routes to generate SNAs \cite{venk3,kt,Prasad1}. The mechanisms for the onset of SNAs are usually classified into three types as: (i) Heagy-Hammel route - the SNAs emerges during the collision of a period doubled torus with its own unstable parent \cite{heag1},  (ii) fractalization route - the truncated torus gets wrinkled and forms SNAs without any interaction with the parent torus \cite{nish}, and (iii) type-III intermittency route - SNAs occur when the torus doubling bifurcation is controlled by sub-harmonic bifurcations \cite{duff1}.  Another possibility for the occurrence of SNAs is through crisis-induced intermittency, wherein the collision of the wrinkled torus with the boundary results in sudden widening of the attractor \cite{venk2}.

\par In this letter, we report the first experimental evidence of SNAs in a self-excited laboratory system, in the absence of external quasiperiodic forcing. We show that the natural dynamics of a laboratory-scale thermoacoustic system \cite{Kabiraj} displays the presence of SNAs in between the quasiperiodic and chaotic attractors.

\par In a thermoacoustic system, the presence of flame in a confined environment at certain conditions leads to the generation of large amplitude, self-sustained tonal sound in the air column of the system, originally known as `singing flame' \cite{Jones} or more recently as `thermoacoustic instability' \cite{Juni}. The occurrence of such self-excited oscillations is detrimental to the structural integrity of practical combustion systems used in propulsion and power generating units \cite{Lieuwen,putnam}. Through the implementation of various time series analysis techniques based on Fourier amplitude spectrum, singular continuous spectrum, spectral distribution function, and $0 - 1$ test, we confirm the presence of SNAs in our system. 

\begin{figure*}[t]
\centering\includegraphics[width=0.75\linewidth]{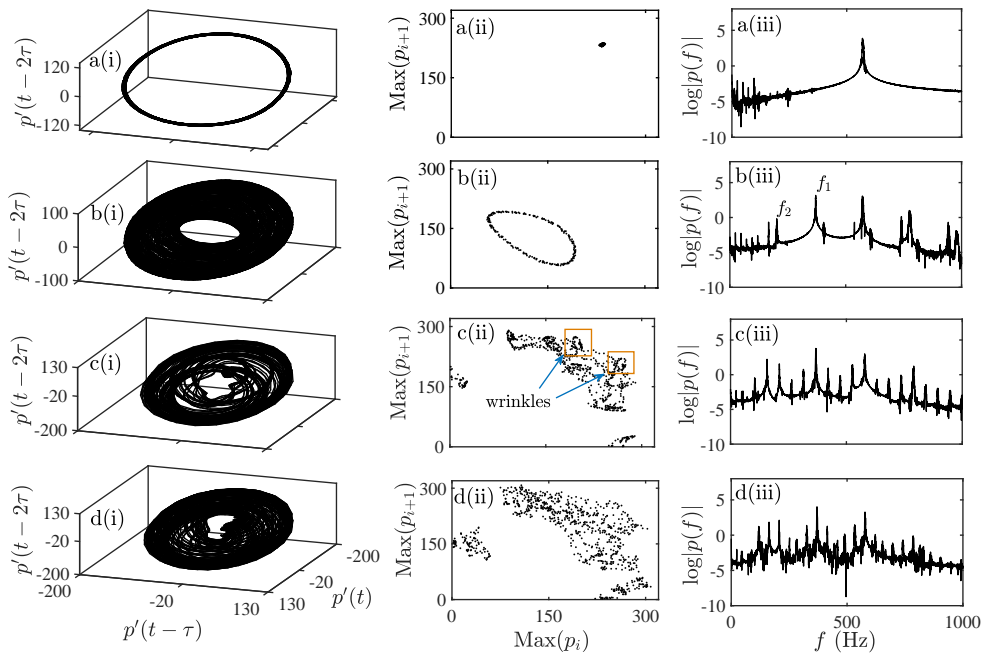}
\caption{(a-d) The dynamics of acoustic pressure obtained at different $x_{f}=0.1725$, $0.195$, $0.22$, and $0.2262$ for the states of limit cycle (P1), quasiperiodicity, SNAs, and chaotic dynamics, respectively. For each state, (i) the phase portrait, (ii) corresponding Poincar\'e section, and (iii) power spectrum are plotted.}
\label{fig:1}
\end{figure*} 
\par The experiments were performed on a laboratory scale ducted laminar premixed flame combustor. More details on the description of the experimental setup, data acquisition, and measurement uncertainties can be found in Kabiraj {\em et al.} \cite{Kabiraj}. The combustor comprises of a transparent borosilicate glass duct of inner diameter $5.67$~cm and length $80$~cm. The glass duct is closed at the bottom end and open at the top end. It consists of a burner tube of length $80$~cm, inner diameter $1.6$~cm, and thickness $0.15$~cm, which is used to supply premixed air and fuel (Liquefied Petroleum Gas) mixture required for combustion.  A circular copper block of $1.8$~cm height, with seven $0.2$~cm sized through holes, is fixed over the burner tube to stabilize the conical flames in the combustor. The equivalence ratio, a measure of relative proportion of the air-fuel mixture involved in the combustion with respect to what is needed for stoichiometric combustor, is fixed at $0.48$ throughout the study. The combustion mixture is ignited from the top of the burner tube using a Butane torch until all flames are stabilized on the copper block.

\par The location of these flames with respect to the open end of the glass duct ($L_f$) is varied as a control parameter in this study. We normalize $L_f$ with the length of the duct ($L$) as $x_{f}=L_f/L$, where $L=80$~cm. In order to change the flame location, the glass duct is moved upwards using a traverse mechanism against the fixed location of the burner tube. The dynamics of the combustor for a given change in the flame location is acquired in terms of acoustic pressure measurements, which are performed using a pressure transducer (PCB 103B02 of sensitivity $= 223.4$ ~mV/kPa and uncertainty = $\pm 0.14$ ~Pa) located $5$~cm from the bottom of the glass duct. The data are acquired using an analog to digital conversion card (NI-6143, 16-bit, resolution = $0.15$ ~mV, voltage range = $\pm 5$~V). The data was acquired for 30 s at the sampling frequency of 10 kHz. The frequency resolution of the power spectrum is 0.03 Hz.

\par Primarily, to understand the dynamical transitions of the considered system and to distinguish each dynamical behavior, we plot the phase portrait [Fig. \ref{fig:1}(a-d)(i)], Poincar\'e section [Fig. \ref{fig:1}(a-d)(ii)], and the corresponding power spectrum [Fig. \ref{fig:1}(a-d)(iii)] of each attractor. With the variation of the flame location ($x_{f}$) in the combustor, we show that the system exhibits a transition from  regular period-1 (P1) oscillation to chaotic dynamics via quasiperiodicity and SNAs. When the flame location is at $x_f=0.1725$, we notice the occurrence of a P1 limit cycle attractor [Fig. \ref{fig:1}(a)]. During this state, the system dynamics evolves on a single periodic orbit in the phase portrait [Fig. \ref{fig:1}a(i)] and shows an isolated point in the Poincar\'e section [Fig. \ref{fig:1}a(ii)]. Further, the presence of a single dominant frequency peak in the power spectrum affirms the P1 nature of the limit cycle attractor [see Fig. \ref{fig:1}a(iii)].

\par For $x_f=0.195$, we notice the existence of quasiperiodic oscillations as a consequence of the interplay between incommensurate frequencies $f_1=369.9$~Hz and $f_2=202$~Hz, and the peaks at their linear combinations observed in the power spectrum [see Fig. \ref{fig:1}b(iii)]. The reconstructed phase space of a quasiperiodic attractor shows a 2-torus structure and a closed loop of points in the Poincar\'e section [Fig. \ref{fig:1}b(i) \& b(ii), respectively]. 

\begin{figure*}[ht]
\centering\includegraphics[width=0.7\linewidth]{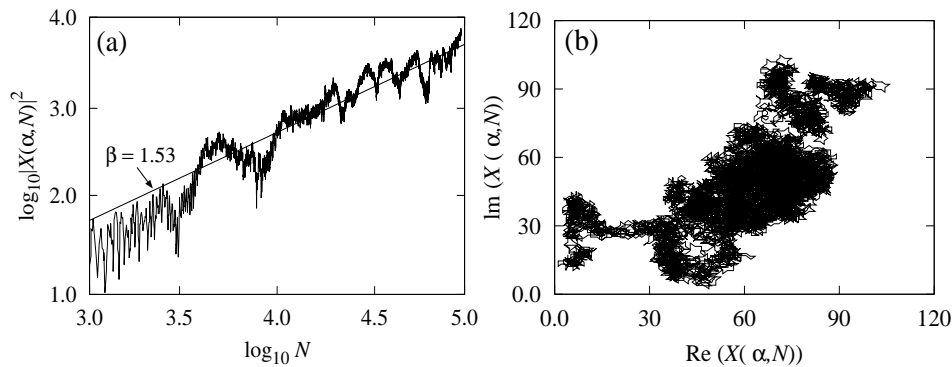}
\caption{Singular continuous spectrum analysis of the acoustic pressure signal obtained at $x_f$ $=$ $0.22$ for SNA. (a) The logarithmic plot of $|X(\alpha,N)|^2$ with respect to $N$ showing the power law scaling, and (b) the fractal path in the complex plane of $X$ confirms the presence of SNAs. The value of $\alpha$ is chosen as 368.4 Hz.}
\label{fig:3}
\end{figure*} 

\par
\begin{figure}[ht!]
	\centering\includegraphics[width=0.7\linewidth]{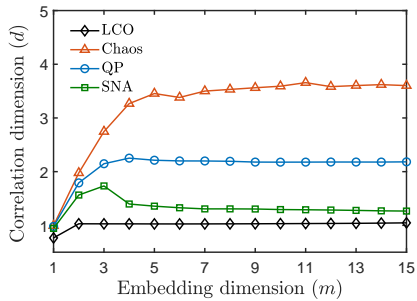}
	\caption{Estimation of correlation dimension ($D$) with respect to embedding dimension ($m$) from the observed time series of different dynamical states such as limit cycle (LCO), quasi-periodicity (QP), strange non-chaos attractor (SNA), and chaos. The unfilled diamond, triangle, circle and square points   denote the LCO, Chaos,  QP, and SNA respectively. }
	\label{cd}
\end{figure}

\begin{figure}[ht!]
	\centering\includegraphics[width=0.99\linewidth]{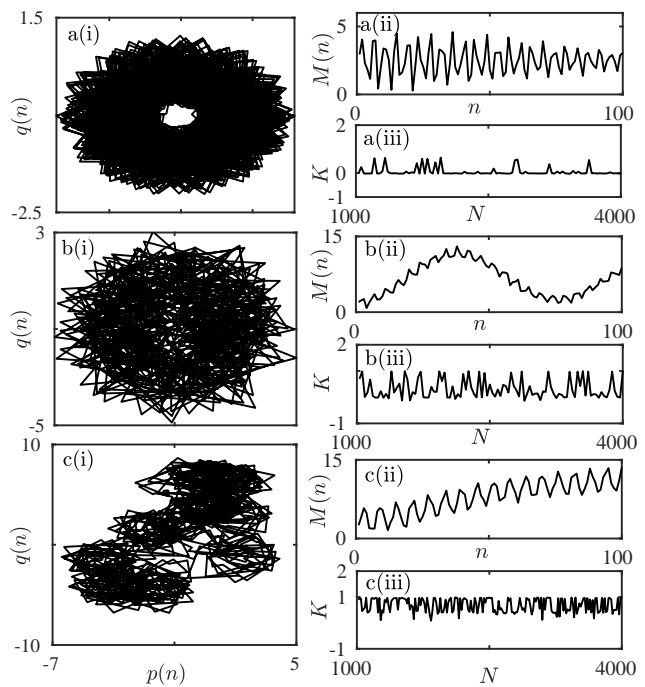}
	\caption{A $0 -1$ test performed to distinguish the dynamics of (a) quasiperiodicity at $x_{f} = 0.1725$, (b) strange nonchaos at $x_{f} = 0.22$, and (c) chaos at $x_{f} = 0.2262$. (i) The plot between the translation variables $p(n)$ and $q(n)$, (ii) the behaviour of mean square displacement $M(n)$ with $n$, and (iii) the variation of growth rate $K$ for these dynamical states.}
\label{fig:5}
\end{figure}

 By changing the flame location to $x_f=0.22$, we find that the stable three-dimensional torus observed for quasiperiodic oscillations [Fig. \ref{fig:1}b(i)] wrinkles and fractalizes [Fig. \ref{fig:1}c(i)] to a strange nonchaotic attractor. This can be seen from the Poincar\'e sections of these attractors, where a clean closed loop structure of quasiperiodic attractor [Fig. \ref{fig:1}b(ii)] breaks down into a wrinkled torus structure of SNA [Fig. \ref{fig:1}c(ii)]. This transition to SNA from quasiperiodicity possibly follows a {\it fractalization} route \cite{nish}; nevertheless, a rigorous confirmation of SNA in the system will be done in Figs. \ref{fig:3}-\ref{fig:5}. The dynamical mechanism beneath this route involves the destabilization and  fractalization of the orbits on the torus without colliding with its parent unstable tori. Hence, the shape and the structure of the parent torus will be retained in this route even when the attractor is transformed into a strange attractor [compare Fig. \ref{fig:1}b(i) \& c(i)]. In addition, the power spectrum in Fig. \ref{fig:1}c(iii) for SNA depicts the presence of many peaks at irrational frequencies, whose relation is not well defined, as witnessed for the case of the quasiperiodic oscillations. Thus, the spectrum with features of neither broadband frequencies (seen for chaotic attractor) nor discrete frequencies (a property of periodic or quasiperiodic attractors) indicates the feature of SNA \cite{paul}.

\par When the flame location is at $x_f=0.2262$, the acoustic pressure inside the combustor exhibits nearly irregular fluctuations, which occurs when the unstable periodic orbits on the SNAs get more destabilized and wrinkled to form a chaotic attractor, as seen from Fig. \ref{fig:1}d(i). The Poincar\'e section of this state shows a scatter of points on its surface [Fig. \ref{fig:1}d(ii)] and the power spectrum indicates a broadband behaviour [Fig. \ref{fig:1}d(iii)], confirming the presence of chaotic dynamics in the signal. From the observed results, it is clear that the transition from periodic to chaotic attractor happens via quasiperiodic and strange nonchaotic attractor in our system.  

\par In the subsequent portion of the paper, we use various tools to qualitatively and quantitatively characterize the dynamics of different dynamical states and also reaffirm the existence of SNAs in our system. In general, a dynamical system can manifest two types of power spectrum namely, continuous and discrete spectrum. Discrete spectrum corresponds to the occurrence of oscillations at specific frequencies such as that of periodic or quasiperiodic oscillations. In contrast, the broadband nature of the frequency spectrum in the case of chaotic oscillations points towards continuous spectrum. For a special case like SNA, the spectrum exhibits a combination of both continuous and discrete components, known as singular continuous spectrum \cite{Yalnkaya,Eckmann,ark1,ark}. In order to calculate the finite time Fourier transform of the experimentally observed time series $x_{k}$, we define the following,
\begin{equation}
X(\alpha,N)=\sum_{k=1}^{N}(x_{k})\exp(2\pi ik \alpha),
\end{equation}
where $\alpha$ corresponds to frequency and $N$ indicates time. Since $X(\alpha, N)$ is a complex variable, the plot between $Re(X)$ and $Im(X)$ helps us understand different dynamical features exhibited by the signal. For regular signals, as the spectrum is discrete, the power $|X(\alpha,N)|^2$ is proportional to $N^2$ and a path on the $(Re(X), Im(X))$-plane displays a persistent and bounded behaviour \cite{ark1}. If the path in the complex plane is random (Brownian walk), then the power of the signal $|X(\alpha,N)|^2$ is directly proportional to $N$, denoting the continuous spectral components observed for chaotic signals. For the singular continuous spectrum, the power $|X(\alpha,N)|^2$ is proportional to $N^{\beta}$ where the value of $1<\beta<2$, the signal posses the properties of SNA \cite{ark1}. The path for SNA on ($Re(X), Im(X)$)-plane adopts a self-similar fractal structure  which is shown in the Appendix II \cite{ark1}. 



To confirm the observed dynamical behaviors, we have estimated the correlation dimension  for each dynamical behavior.  To accomplish this measure, we first reconstruct the data as interms of embedded delay,   
$X(t, \tau)=[x(t); x(t-\tau); X(t-2\tau;  ...x(t-m \tau)]$,   
where, $\tau$ is the time delay and m is the embedding dimension. Then the correlation dimension $D$ can be obtained using the expression, 
\begin{eqnarray}
D= \lim_{r\rightarrow 0} \frac{ln~ C(\epsilon)}{ln~ \epsilon}
\end{eqnarray}
Where, $\epsilon$ is the distance and  $C(\epsilon)$ is the correlation sum is identified using the relation,
\begin{eqnarray}
C(\epsilon) = \lim_{N \to \infty}\frac{1}{N} \sum_{i,j}^N \Theta(\epsilon-|x_i-x_j|),   
\end{eqnarray}
Where $\Theta$ is the Heaviside function. If the distance $\epsilon$ is positive, then $\Theta(\epsilon)=1$ else $\Theta(\epsilon)=0$. $x_i$ and $x_j$ are the locations of two trajectory points. The value of $C(\epsilon)$ changes as the distance $\epsilon$ decreases and obeys the power-law,  $C(\epsilon)=\epsilon^D$. 
To validate the observed dynamical, we have plotted the correlation dimension ($D$) as a function of embedding dimension ($m$) in Fig.~\ref{cd}. Each dynamical behavior represented line connecting by diamond, circle, triangular, and square points correspond to the limit cycle oscillations (LCO), quasi-periodic (QP), strange non-chaotic (SNA), and Chaos, respectively.    For periodic limit cycle oscillation, the correlation dimension is saturated at the critical value $D=1$.  For the quasi-periodic oscillations and SNA, D  value  saturates at $2.1$ and $1.3$ which is also confirms the emergence of quasiperiodic and strange nonchaotic evolution.  The correlation dimension greater than $3$ indicates the chaotic behavior of the system. This quantitative measure substantiate the presence of, periodic, quasi-periodic, SNA and chaotic behaviors in the chosen experimental setup.

\par In order to distinguish the dynamics of SNA from chaos, we make use of the $0-1$ test \cite{gott}, as suggested by Gopal {\em et al.} \cite{Gopal}. The usual practice to distinguish these dynamical states is to calculate the maximum Lyapunov exponent of the signal, whose value is positive for a chaotic signal and negative for SNA \cite{Kantz}. However, as most of the experimental data contains intrinsic noise, the computation of maximum Lyapunov exponent gets challenging. Hence, confirming chaos in the system dynamics using Lyapunov exponent becomes unfeasible \cite{feud}.
\par The first step in implementing the $0-1$ test is to compute the translation variables, denoted as $p(n)$ and $q(n)$, from the input time series $\phi(j)$ such that \cite{gott},
\begin{equation}
p(n) = \sum_{j=1}^n \phi(j)\cos(jc) 
~ ~ and ~ ~ q(n) = \sum_{j=1}^n \phi(j)\sin(jc).
\end{equation}

Here, $n=1, 2,..., N$. The value of the constant $c$ can be chosen in the interval $(\pi/5, 4\pi/5)$. The behaviour of these two new variables helps in distinguishing different dynamical states in the system. For regular dynamics (periodic or quasiperiodic), the behaviour of these variables is bounded, while it is unbounded or drifting for the chaotic dynamics. The motion of translation variables depends on the value of $n$, which is much less than $N$ and often chosen as $n=N/10$ \cite{gott}. The behaviour of the trajectory in the ($p(n), q(n)$)-plane for increasing $n$ can be calculated through the mean square displacement $D(n)$ as follows, 
\begin{equation}
D(n) = \frac{1}{N}\sum_{j=1}^n([p(j+n)-p(j)]^2 + [q(j+n)-q(j)]^2).
\end{equation} 
In order to resolve convergence issues of $D(n)$, a modified mean square displacement is used \cite{gott,gott1}, which is obtained as follows
\begin{equation}
M(n)=D(n)-V_{osc}(c,n), 
\end{equation} 
where $V_{osc}(c,n)$ $=$ $\frac{1}{N} \sum_{j=1}^n \phi(j) \frac{(1-\cos (jc))}{(1-\cos (c))}$.
For a chaotic signal, the value of $M(n)$ will linearly increase with $n$; whereas, for a regular signal, it remains nearly constant \cite{gott,gott1}.
Further, the asymptotic growth rate of such mean displacements is calculated through a linear regression, which is given by the following equation,
\begin{equation}
K = \lim_{n \rightarrow \infty}  \frac{\log M(n)}{\log n}.
\end{equation}     
The value of $K$ lies between $0$ and $1$ \cite{nicol}.  If the dynamics is chaotic, $K$ takes a value close to $1$, and for a regular signal it approaches  $0$. In order to distinguish dynamics of SNA from regular and chaotic oscillations, Gopal {\em et al.} \cite{Gopal} suggested that the choice of $c$ should be the Golden mean ratio, i.e., $c=(\sqrt 5+1)/2$. For SNAs, the value of $K$ lies between $0$ and $1$.
\par In Fig. \ref{fig:5}(a-c), we plot the translation variable $p(n)$ versus $q(n)$ for the dynamics of quasiperiodic, SNA, and chaotic oscillations, respectively. The trajectory in $(p, q)$-plane appear to be bounded along a circle, indicative of quasiperiodic dynamics [Fig. \ref{fig:5}(a)] in the pressure signal obtained at $x_{f}=0.1725$. Furthermore, the mean displacement $M(n)$ exhibits fluctuations around some constant value [Fig. \ref{fig:5}a(ii)] and the growth rate shows a value near zero [Fig. \ref{fig:5}a(iii)], confirming the quasiperiodic dynamics \cite{Gopal} of the pressure oscillation observed during this state. On the other hand, for the chaotic signal [Fig. \ref{fig:5}c(i)], the trajectory in $(p, q)$-plane show a random-walk (or Brownian) type behaviour. The variation of $M(n)$ with $n$ displays an increasing trend [Fig. \ref{fig:5}c(ii)] with a growth value ($K$) near unity [Fig. \ref{fig:5}c(iii)], further affirming the presence of chaotic oscillations in the pressure signal obtained at $x_f=0.2262$. Since, SNA consists of properties of both regular and chaotic dynamics, we notice the presence of bounded trajectory with a minimal Brownian structure \cite{Gopal} in $(p, q)$-plane [Fig. \ref{fig:5}b(i)]. The presence of SNAs can also be confirmed from the plot of variation of $M(n)$ with $n$, where the mean of this plot does not increase monotonically but shows an oscillatory behaviour with $n$, along with the value of $K$ lying between $0$ and $1$ \cite{gott,gott1}. Thus, using the $0-1$ test, we distinguish the features of SNA from quasiperiodic and chaotic dynamics in our system.

\par In summary, we report the first experimental evidence of SNAs in the natural dynamics of a laboratory-system which is not quasiperiodically forced. This observation is in contrast to the usual experimental studies on SNAs that requires quasiperiodic forcing for the birth of SNAs. We witness the existence of SNAs in between the states of quasiperiodic and chaotic dynamics in a laminar thermoacoustic system when the flame location in the combustor is varied as the control parameter. The presence of SNAs is confirmed through various characterization tools such as singular-continuous spectrum, spectral distribution function, and $0-1$ test. The birth of SNAs is shown to happen via fractalization route. 

In general, the observation of SNAs through experiments in unforced practical systems is a herculean task, as such attractors occur only in a narrow interval of control parameter between quasiperiodicity and chaos. Therefore, for the experimental realization of the SNAs, most researchers have to rely on the need of quasiperiodic forcing in the system \cite{Guan,kt,venk2,prem,paul}. The presence of SNA in the system dynamics has been projected to have wide applications ranging from a secure communication, ease of synchronization to computation process \cite{Prasad1}. However, the implementation of SNAs in real-time applications is still a topic of investigation. The experimental realization of SNAs in self-excited dynamics of a laboratory system is a first step in realizing the possibility of such dynamics in practical systems without any forcing. We believe that the existence of SNAs would be more ubiquitous in self-excited systems than previously thought. 

\par We gratefully acknowledge J. C. Bose fellowship (JCB/2018/000034/SSC) and Swarnajayanti fellowship (DST/SF/1(EC)/2006) from Department of Science and Technology (DST), Government of India for the financial support.


\end{document}